\documentclass[preprint]{aastex631}
\usepackage{amsmath}
\usepackage{graphicx}

\submitjournal{\apj}

\accepted{\today}

\shorttitle{Slow Shock Instability v.s. Ambipolar Diffusion}
\shortauthors{Abe et al.}

\begin{document}

\title{Growth of Massive Molecular Cloud Filament by Accretion Flows I: Slow Shock Instability v.s. Ambipolar Diffusion}
\email{d.abe@nagoya-u.jp}

\author[0000-0001-6891-2995]{Daisei Abe}
\affiliation{Department of Physics, Graduate School of Science, Nagoya University, Furo-cho, Chikusa-ku, Nagoya 464-8602, Japan}

\author[0000-0002-7935-8771]{Tsuyoshi Inoue}
\affiliation{Department of Physics, Faculty of Science
and Engineering, Konan University, Okamoto 8-9-1, Higashinada-ku, Kobe 658-8501, Japan}
\affiliation{Department of Physics, Graduate School of Science, Nagoya University, Furo-cho, Chikusa-ku, Nagoya 464-8602, Japan}

\author[0000-0003-4366-6518]{Inutsuka Shu-ichiro}
\affiliation{Department of Physics, Graduate School of Science, Nagoya University, Furo-cho, Chikusa-ku, Nagoya 464-8602, Japan}

\graphicspath{{./}{figures/}}

\begin{abstract}
{The \textit{Herschel} Gould Belt Survey showed that stars form in dense filaments in nearby molecular clouds.}
Recent studies suggest that massive filaments are bound by the slow shocks caused by accretion flows onto the filaments.
{The slow shock is known to be unstable to corrugation deformation of the shock front. The corrugation instability could convert the accretion flow’s ram pressure into turbulent pressure {that influences the width of the filament, which, according to theory, determines the self-gravitational fragmentation scale and core mass.} In spite of its importance, the effect of slow shock instability on star-forming filaments has not been investigated. In addition, the linear dispersion relation obtained from the ideal magnetohydrodynamics (MHD) analysis shows that the most unstable wavelength of shock corrugation is infinitesimally small (or mean free path).}
In the scale of dense filaments, the effect of ambipolar diffusion can suppress the instability at small scales.
This study investigates the influence of ambipolar diffusion on the instability of the slow shock.
We perform two-dimensional MHD simulations to examine the linear growth of the slow shock instability, considering the effect of ambipolar diffusion.
The results demonstrate that the most unstable scale of slow shock instability is approximately five times the length scale of ambipolar diffusion $\ell_{\mathrm AD}$ calculated using post-shock variables, {where, $\ell_{\mathrm AD}$ corresponds to the scale where the magnetic Reynolds number for ambipolar diffusivity is unity.}
\end{abstract}


\section{Introduction} \label{sec:intro}
The dense filamentary structures in {nearby} molecular clouds are sites of star formation~\citep[e.g.,][]{andre2010A&A...518L.102A,Hacar2022arXiv220309562H}.
{The} \textit{Herschel} Gould Belt survey reported that stars are formed in filaments with {line-masses over which gravity wins over thermal pressure}, $M_{\mathrm{line,cr,th}} = 2c_{\mathrm{s}}^2 /G \simeq 17\ \mathrm{M_{\odot}\ pc^{-1}},$ where, $c_{\mathrm{s}} \simeq 0.2\ \mathrm{km\ s^{-1}}$ and $G$ denote the isothermal sound speed of typical molecular clouds and gravitational constant, respectively~\citep[e.g.,][]{Stodolkiewicz1963,Ostriker1964,InutsukaMiyama1992,Inutsuka1997}.
Several authors have studied the formation mechanism of filaments~\citep[e.g., ][]{Tomisaka1983,Nagai1998,PadoanNordlund1999ApJ...526..279P,Hennebelle2013,Pudritz2013RSPTA.37120248P,inoue2013ApJ...774L..31I,chenOstriker2014ApJ...785...69C,inutsuka2015A&A...580A..49I,Balfour2017,Federrath2016,abe2021ApJ...916...83A}.
Recently, \citet{abe2021ApJ...916...83A} classified the proposed formation mechanisms into Type G, C, O, I, and S, which are summarized in Table \ref{tab:FormationMechanism}.
\begin{table*}
\caption{Filament formation mechanisms.\label{tab:FormationMechanism}}
 \centering
  \begin{tabular}{ccl}
   \hline
   Category & Filament vs. Magnetic field & A brief description of the formation mechanism\\
   \hline \hline
    Type G & perpendicular & Sheet-like clouds fragment into filaments by self-gravity. \\
    & & \\
    \hline
    Type I & - & Filaments arise at the intersection line between two   \\
    & & shock-compressed sheets. \\
    \hline
    Type O & perpendicular & Filaments form at the convergent point of gas flows  \\
    & & within deformed oblique MHD shock fronts induced\\
    & & by the clumpiness of the medium.\\
    \hline
    Type C & perpendicular & Gas coagulation along the magnetic field by local turbulent velocity  \\
    & &  perturbations within shock-compressed layers.\\
    \hline
    Type S & parallel & Shear flows associated with turbulence stretch existing clumps. \\
   \hline
  \end{tabular}
\end{table*}
{In general, mechanisms G, C, and O result in supercritical filaments, which are of interest for star formation.}
The common feature of these mechanisms is that filaments are formed by gas flow along the local magnetic field in a shocked-compressed sheet.

{Several recent works highlight the importance of accretion in the context of a filament's evolution.}
{For example,} molecular emission-line observations provide evidence of the perpendicular accretion onto filaments~\citep{Palmeirim2013,shimajiri2019A&A...632A..83S,Chen2020}.
In particular, \citet{shimajiri2019A&A...632A..83S} reported the occurrence of accretion onto filaments in a shocked sheet.
{\citet{Clarke2016MNRAS.458..319C} demonstrated that the most unstable length scale for self-gravitational fragmentation along a filament depends on the accretion rate onto the filament.
\citet{HennebelleAndre2013A&A...560A..68H} developed an analytical model that can be applied to self-gravitating and accreting filaments.
They considered turbulence driven by accretion onto the filament and its dissipation by the ion-neutral friction.}

\begin{figure}[ht!]
\epsscale{0.5}
\plotone{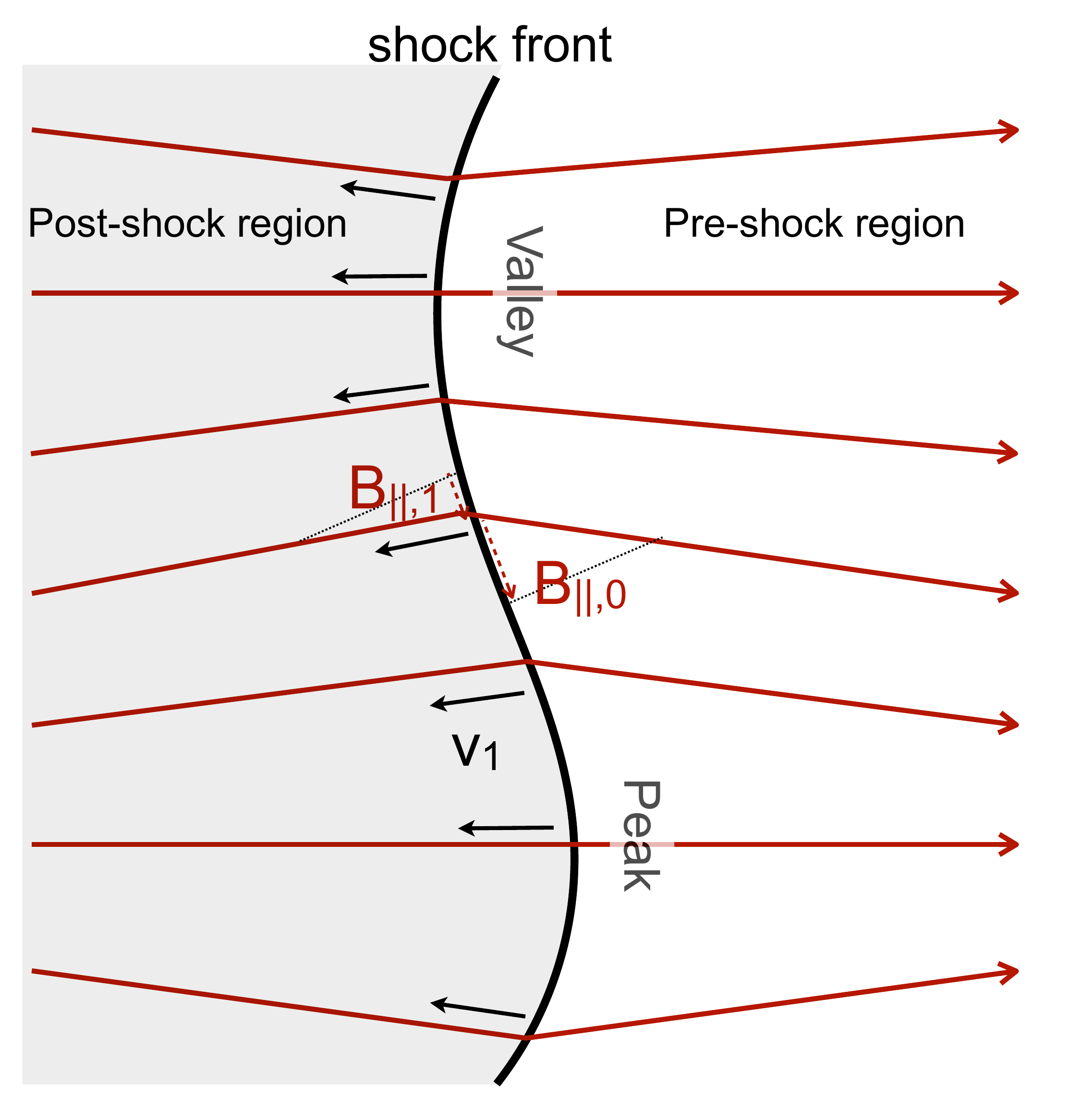}
\caption{Schematic illustration of two-dimensional slow magnetohydrodynamic shock.\label{fig:ss}
}
\end{figure}
A shock wave with Alfv\'{e}n Mach number $\mathcal{M}_{\rm A}<1$ and sonic Mach number $\mathcal{M}_{\rm s}>1$ is called ``slow (mode) shock."
{Given that massive filaments are formed in the post-shock layer threaded by a strong magnetic field with energy exceeding the kinetic energy of accretion flows~\citep[Type O mechanism, ][]{Inoue2018PASJ...70S..53I}, the filament surface is naturally bound by the slow shocks.}
\citet{LessenDeshpande1967JPlPh...1..463L} found through linear stability analysis that the slow shock front is corrugationally unstable (slow shock instability, hereafter SSI).
The mechanism of SSI is expressed as follows.
In contrast to the fast shock, the component of the magnetic field tangential to the shock surface decreases across the front.
Thus, when the shock corrugates, the magnetic field lines kink as denoted by red lines in Figure \ref{fig:ss}.
Because gas flows along the magnetic field, the gas converges behind the peak of the shock, while it diverges behind the valley.
Such flow patterns increase (decrease) the pressure behind the peaks (valleys), which further push up (pull down) the shock front.
\citet[][]{Edelman1989Ap.....31..758E} showed that the approximated dispersion relation of SSI for $\mathcal{M}_{\rm A}\ll 1$ can be written as
\begin{equation}
\omega \sim -i \frac{\gamma-1}{\gamma+1} M_{\mathrm{A}}^2 v_{\rm sh} k,
\label{eqs:ana disp arbitrary gamma}
\end{equation}
where, $\omega$, $v_{\mathrm{sh}}$, and $k$ denote the frequency, shock velocity, and wave number of the shock corrugation.
As an more accurate solution, \citet[][]{Edelman1989Ap.....31..758E} derived the approximate dispersion relation for $\gamma=5/3$:
\begin{equation}
\omega\simeq-i\left(0.208 M_{\mathrm{A} }^{2}-0.0775 M_{\mathrm{A}}^{4}+0.06 M_{\mathrm{A} }^{6}\right) v_{\mathrm{sh}} k.
\label{eqs:ana disp}
\end{equation}
Eqs. (\ref{eqs:ana disp arbitrary gamma}) and (\ref{eqs:ana disp}) indicate that the most unstable scale is infinitesimally small. This unphysical feature stems from the ideal approximation and the resulting discontinuous treatment of the shock.
To know the physical scale length of the SSI, we consider a non-ideal effect.
Since the corrugation of the shock front generally produces turbulent flows behind the shock \citep[e.g.,][]{Inoue2012ApJ...744...71I,InoueInutsuka2012ApJ...759...35I}, we can expect that the SSI will deposit additional energy to the filament.

In molecular clouds, the ambipolar diffusion is effective and potentially modifies the SSI dynamics.
The magnetic Reynolds number of the flow with ambipolar diffusivity is given as
\begin{equation}
\mathcal{R}_{\mathrm{AD}}= \frac{4\pi \gamma_{\mathrm{in}} \rho_{\mathrm{n}}\rho_{\mathrm{i}} v\ell}{ B^2},
\end{equation}
where, $\gamma_{\mathrm{in}} \equiv \langle \sigma_{\mathrm{in}}v_{\rm in} \rangle / (m+m_{\mathrm{i}}) = 3.5\times10^{13}\ \mathrm{cm^3\ g^{-1}\ s^{-1}}$ and $\rho_{\mathrm{i}}$ denotes the ion mass density.
$\sigma_{\mathrm{in}}$, $v_{\rm in}$, $m$, and $m_{\mathrm{i}}$ represent {the ion-neutral cross-section (Langevin cross-section)}, the relative velocity between a neutral molecule and ion, mean molecule mass, and mean ion mass, respectively.
Assuming a balance between the ionization by cosmic rays and the recombination, $\rho_{\mathrm{i}}$ can be expressed as $C \rho^{1/2}$.
In this study we apply $C=3\times 10^{-16}$ cm$^{-3/2}$ g$^{1/2}$~\citep[][]{shu1992pavi.book.....S}.
The characteristic length scale below which the effect of ambipolar diffusion becomes non-negligible can be obtained by solving $\mathcal{R}_{\mathrm{AD}}= 1$ that yields
\begin{equation}
\ell_{\mathrm{AD}}= 0.09\ \mathrm{pc}\ \left(\frac{B}{30\ \mathrm{\mu G}}\right)^{2}\left(\frac{n}{10^3\ \mathrm{cm^{-3}}}\right)^{-3/2}\left(\frac{v}{1\ \mathrm{km/s}}\right)^{-1},
\label{eqs:ambi scale}
\end{equation}
Comparing the actual observation, for densities around 5,000 cm$^{-3}$, the mean magnetic field from the \citet{Crutcher2012} plot (Fig. 6 in that review) is about 5 $\mu$G.
Then Eq. \ref{eqs:ambi scale} gives $\ell_{\mathrm{AD}}\sim 2.2\times 10^{-4}$ pc.
The observed maximum magnetic field strength for the same density range is about 50 $\mu$G, for which we get $\ell_{\mathrm{AD}}\sim 2.2 \times 10^{-2}$ pc.
Therefore, $\ell_{\mathrm{AD}}$ can take a wide range of values.
{This suggests that the characteristic scale of the ambipolar diffusion can be comparable to the filament width and hence the ambipolar diffusion can affect the filament dynamics.}
In the context of solar chromosphere, \citet[][]{SnowHillier2021MNRAS.506.1334S} performed two-dimensional two-fluid simulations of SSI for partially ionized gas.
They demonstrated that the neutral fluid stabilizes the SSI on a small scale and found new features such as gas accumulation at valleys.
However, the situation in their simulations is different from the one in molecular clouds (e.g., ionization degree, ion-neutral collision cross-section, etc.), and they did not study the dispersion relation and the dependence on density and the magnetic field.
The linear analysis of SSI including ambipolar diffusion is challenging.
Our strategy is to directly simulate the SSI including the ambipolar diffusion and the measurement of the growth rate.

In this paper, {as a first step to understand the effect of SSI on filaments}, we study the effect of ambipolar diffusion on the SSI and derive the most unstable scale.
{As a result of this study, we can determine the typical length scale of SSI in filaments that provides the resolution requirement in future simulations.}
More realistic simulations of filament evolution with slow shocks caused by the converging accretion including self-gravity will be our future studies.
The paper is organized as follows: In \S \ref{sec:setup}, we provide the setup of our simulations, and we show and interpret the results in \S \ref{sec:Results}.
{In \S \ref{sec: Discussion}, we discuss the stabilizing scale of SSI versus ambipolar diffusion.
Finally, we summarize the results in \S \ref{sec: summary}.}

\section{Setup for simulations} \label{sec:setup}

We perform two-dimensional {and three-dimensional} ideal/non-ideal MHD simulations using Athena++ code~\citep{stone2020ApJS..249....4S}.
{To determine the physical scale of the slow shock instability, 2D simulation seems to be sufficient, because the linear stability analysis do not show the difference. In addition, by ideal MHD simulations, \citet{StoneEdelman1995ApJ...454..182S} demonstrated that the growth rates of the slow shock instability in three-dimensional cases are not different from those in two-dimensional cases. We confirm this expectation even with the effect of AD in \S\ref{subsubsec: Three dimensional simulations}.}
We use the second-order accurate van Leer predictor-corrector scheme and piecewise linear method applied to primitive variables to integrating the equations.
The constrained transport method~\citep{StoneGardiner2009NewA...14..139S} ensures the divergence-free condition, $\nabla \cdot \boldsymbol{B} =0$.
In this paper, we do not solve the Poisson equation for self-gravity because we concentrate on the physics of SSI under the influence of ambipolar diffusion as the first step of this sort of study.
The effect of self-gravity will be considered in our future studies.
We solve the following equations:
\begin{equation}
\frac{\partial \rho}{\partial t}+\nabla \cdot(\rho \boldsymbol{v})=0
\label{eqs:eoc}
\end{equation}
\begin{equation}
\frac{\partial \rho \boldsymbol{v}}{\partial t}+\nabla \cdot\left(\rho \boldsymbol{vv}-\frac{\boldsymbol{B B}}{4\pi}+P^{*}\boldsymbol{I}+\boldsymbol{\Pi}\right)=0
\label{eqs:eom}
\end{equation}
\begin{eqnarray}
\frac{\partial E}{\partial t}
+\nabla \cdot \left[ \left(E+P^{*}\right) \boldsymbol{v}
-\boldsymbol{B}(\boldsymbol{B} \cdot \boldsymbol{v}) 
+\boldsymbol{\Pi} \cdot \boldsymbol{v} + \frac{\eta_{\mathrm{AD}}}{|\boldsymbol{B}|^{2}}\{\boldsymbol{B} \times(\boldsymbol{J} \times \boldsymbol{B})\} \times \boldsymbol{B} \right]=0
\label{eqs:ee}
\end{eqnarray}
\begin{equation}
\frac{\partial \boldsymbol{B}}{\partial t} - \boldsymbol{\nabla} \times \left[(\boldsymbol{v} \times \boldsymbol{B}) - \frac{\eta_{\mathrm{AD}}}{|\boldsymbol{B}|^{2}} \boldsymbol{B} \times(\boldsymbol{J} \times \boldsymbol{B})\right]=0,
\label{eqs:ie}
\end{equation}
where, $P^{*}=p+B^{2} / (8\pi)$ and $E=e+ \rho v^{2}/2 + B^{2}/(8\pi)$ denote the total pressure and total energy density; $\rho, p, \boldsymbol{v},$ and $\boldsymbol{B}$ represent the density, pressure, velocity, and the magnetic field; $\boldsymbol{J}=\nabla \times B$ represents the current.
We introduce the viscous stress tensor
\begin{equation}
\Pi_{i j}=\rho \nu\left(\frac{\partial v_{i}}{\partial x_{j}}+\frac{\partial v_{j}}{\partial x_{j}}-\frac{2}{3} \delta_{i j} \nabla \cdot v\right)
\end{equation}
to prevent the carbuncle phenomenon~\citep{Quirk1994IJNMF..18..555Q,Liou2000JCoPh.160..623L,Kim2003JCoPh.185..342K} and the growth of a grid scale SSI seeded by the carbuncle instability.
$\nu$ denotes the coefficient of physical kinematic viscosity, which is adjusted to stabilize a grid scale (eight cells) fluctuation.
The box size $L_{\rm box}$ and $\nu$ are chosen so that the stabilizing scale by ambipolar diffusion is sufficiently {larger} than this grid scale.
$\eta_{\mathrm{AD}}$ denotes the ambipolar diffusion coefficient, which is given by
\begin{equation}
\eta_{\mathrm{AD}} = \frac{B^{2}}{4\pi \gamma_{\mathrm{in}} \rho_{\mathrm{n}} \rho_{\mathrm{i}}},
\end{equation}
{where, $\rho_{\mathrm{n}} \simeq \rho$ represents the neutral gas mass density, and the ion mass density is denoted by $\rho_{\mathrm{i}}=C \rho^{1/2}$. In this study we apply $C=3\times 10^{-16}$ cm$^{-3/2}$ g$^{1/2}$~\citep[][]{shu1992pavi.book.....S}.}

We numerically solve Eqs. (\ref{eqs:eoc})--(\ref{eqs:ie}) on a two-dimensional domain of size [-4$L_{\mathrm{box}}$, 4$L_{\mathrm{box}}$] $\times$ [0 pc, $L_{\mathrm{box}}$] in the shock rest frame.
We select $L_{\mathrm{box}}=0.2$, 0.25, or 0.5 pc.
The specific heat ratio $\gamma=1.01$ is used.
The initial density, velocity, and pressure field is set as
\begin{equation}
\rho(x,y) = \rho_0 \left[ 1 + \frac{r-1}{2}(1 - \tanh[x/0.01 \rm{pc}]) \right] + \rho_{\mathrm{p}},
\end{equation}
\begin{equation}
{v}_{x}(x,y) = -v_{x0}\left[ 1 + \frac{r-1}{2}(1 - \tanh[x/0.01 \rm{pc}]) \right]^{-1},
\end{equation}
and
\begin{equation}
p(x,y) = p_0\left[
1 +
\frac{r_{\mathrm{pres}}-1}{2}
(1-\tanh(x/0.01 \rm{pc}))
\right]
\end{equation}
respectively, where, $\rho_0, v_{x0},$ and $p_0$ denote the initial density, x-component of velocity, and pressure in the pre-shock region, respectively.
We set the upstream gas sound speed $c_{\mathrm{s}}$ as 0.2 km s$^{-1}$ so that the $p_0$ is given by $p_0=\rho_0 c_{\mathrm{s}}^2 / \gamma$.
The compression ratio $r$ and pressure jump $r_{\mathrm{pres}}$ can be written as
\begin{equation}
r \equiv \frac{(\gamma + 1) \mathcal{M}^2_{\mathrm{s}}}{(\gamma - 1)\mathcal{M}^2_{\mathrm{s}}+2},
\end{equation}
and
\begin{equation}
r_{\mathrm{pres}} \equiv \frac{2 \gamma \mathcal{M}^2_{\mathrm{s}} - (\gamma - 1)}{\gamma + 1}.
\end{equation}
As a seed of instability, the density perturbation is introduced as follows.
\begin{equation}
\rho_{\mathrm{p}} = 10^{-4}\times \rho_0 \cos\left(\frac{2\pi y}{\lambda_{\mathrm{p}}}\right) \sin\left(\pi \frac{x}{0.01L_{\mathrm{box}}}\right),\ 
\mathrm{if}\ 0.005L_{\mathrm{box}} \leq x \leq 0.015L_{\mathrm{box}}
\end{equation}
{For a three-dimensional simulation, the density perturbation is}
\begin{equation}
\rho_{\mathrm{p}} = 10^{-4}\times \rho_0 \cos\left(\frac{2\pi y}{\lambda_{\mathrm{p}}}\right) \cos\left(\frac{2\pi z}{\lambda_{\mathrm{p}}}\right) \sin\left(\pi \frac{x}{0.01L_{\mathrm{box}}}\right),\ 
\mathrm{if}\ 0.005L_{\mathrm{box}} \leq x \leq 0.015L_{\mathrm{box}}
\end{equation}
where, $\lambda_{\mathrm{p}}$ denotes the wavelength of perturbation.
These initial conditions lead to a perturbed stationary shock at $x$ = $0$.
Since the star-forming filaments are perpendicular to the magnetic field, the initial uniform magnetic field is set along the x-axis $B_0\hat{x}$.
The numerical domain is a 2D box with a uniform grid of 4096 $\times$ 512 cells, which leads to a spatial resolution of $\Delta x$~=~$L_{\mathrm{box}}$~/~512.
{We apply zero-gradient boundary conditions (with the continuous gas flow) at the boundaries $x=-4L_{\mathrm{box}}$ and $x=4L_{\mathrm{box}}$.}
For $y$ = 0, $L_{\mathrm{box}}$ boundaries, we used the periodic boundary conditions.

We simulate totally 24 different models.
Each model has a unique name, starting with ``n" (for ``upstream density $n_0$"), followed by the number density (``800,'' ``1000," ``1300," and ``1600" [cm$^{-3}$]), the magnetic field (``b"), followed by the field strength (``24," ``30," ``35," and ``40" [$\mu$G]), and the velocity (``v"), followed by the upstream velocity (``0.8," ``0.9," and ``1" [km s$^{-1}$]).
Models with ambipolar diffusion are additionally denoted as ``AD."
The set of parameters used in our simulations are listed in Table \ref{tab:modelparameters}. 
We perform a lot of simulations to test the optimal solver, viscosity, and the super-time stepping method.
The set of parameters are listed in Table in \ref{tab:modelparameters2} in Appendix.

\begin{table*}
\caption{Model parameters.\label{tab:modelparameters}}
 \centering
  \begin{tabular}{ccccccc}
   \hline
  Model Name  & $n_0$  & $B_{0}$ & $v_{\mathrm{x0}}$ & ambipolar & STS & Dimension \\
  & [cm$^{-3}$] & [$\mu$G] & [km s$^{-1}$] & diffusion &\\
   \hline \hline
    n1000b30v1      & 1000 & 30  & 1.0  & No & No & 2D   \\
    n1000b30v1AD    & 1000 & 30  & 1.0  & Yes & No & 2D  \\
    n800b30v1AD     &  800 & 30  & 1.0  & Yes & No & 2D  \\
    n1300b30v1AD    & 1300 & 30  & 1.0  & Yes & No & 2D  \\
    n1600b30v1AD    & 1600 & 30  & 1.0  & Yes & No & 2D  \\
    n1000b24v1AD    & 1000 & 24  & 1.0  & Yes & No & 2D  \\
    n1000b24v1AD3D  & 1000 & 24  & 1.0  & Yes & Yes & 3D  \\
    n1000b35v1AD    & 1000 & 35  & 1.0  & Yes & No & 2D  \\
    n1000b40v1AD    & 1000 & 40  & 1.0  & Yes & No & 2D  \\
    n1000b30v0.8AD  & 1000 & 30  & 0.8  & Yes & No & 2D  \\
    n1000b30v0.9AD  & 1000 & 30  & 0.9  & Yes & No & 2D  \\
    
   \hline
  \end{tabular}
   \tablecomments{We use Roe solver. $\mathcal{R}_{\mathrm{shear}} \equiv v_{\mathrm{x0}} \Delta x/{\nu} = 19.5$ are used (see \S\ref{subsec: SSI in an ideal MHD case} for the selection of $\mathcal{R}_{\mathrm{shear}}$).}
\end{table*}

\section{Results: Dispersion Relation of SSI in Molecular Clouds} \label{sec:Results}

\subsection{SSI in an ideal MHD case} \label{subsec: SSI in an ideal MHD case}
\begin{figure}[ht!]
\plotone{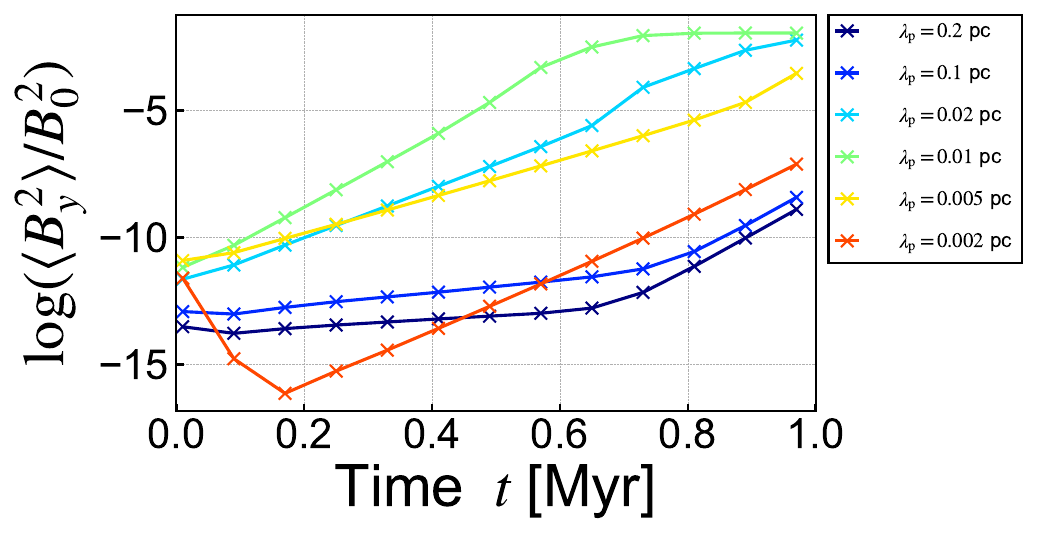}
\caption{\small{Evolution of mean value of the perturbed magnetic field for the case with isothermal ideal MHD including physical shear viscosity (model n1000b30v1).\label{fig:byevo_iso}
}}
\end{figure}

Since the isothermal treatment is justified in dense regions of molecular clouds, we adopt $\gamma = 1.01$.
We use the Roe solver because of its more numerically stable features in nonlinear regimes (see Appendix \ref{subsubsec: Unphysical Numerical Explosion}).
The method to measure the growth rate of the SSI is the same as the method developed by~\citet[][]{StoneEdelman1995ApJ...454..182S}.
It is convenient to use compression-weighted averages because we need to evaluate physical quantities in the vicinity of a shock wave.
The compression-weighted transverse magnetic energy is written as follows.
\begin{equation}
\left\langle B_{y}^{2}\right\rangle=\frac{\int B_{y}^{2} C d V}{\int C d V}
\end{equation}
where,
\begin{equation}
C=\min \left(\partial v_{x} / \partial x, 0\right).
\end{equation}
In Figure \ref{fig:byevo_iso}, we show the evolution of $\left\langle B_{y}^{2}\right\rangle$ in model n1000b30v1.
We can confirm linear growth for $\lambda_{\rm p}$ = 0.005 -- 0.2 pc modes for $t$ $\sim$ 0.1 -- 0.6 Myr.
For $\lambda_{\mathrm{p}}=0.002$ pc, $\left\langle B_{y}^{2}\right\rangle/B^2_0$ decreases until $\sim$ 0.2 Myr, then a larger scale ($>\ 0.002$ pc) grid noise grows after $t \sim$ 0.2 Myr, which is different from the growth of $\lambda_{\mathrm{p}}=0.002$ pc mode of the SSI.
{Also for $\lambda_{\mathrm{p}}$ = 0.2, 0.1, 0.02, and 0.005 pc, we can see the slope increments of the perturbed magnetic field after $t$ = 0.7 Myr caused by grid noise.
(In these cases, the scale of noise is smaller than $\lambda_{\mathrm{p}}$.)}
$\left\langle B_{y}^{2}\right\rangle/B^2_0$ converges to $-4$ after $t = 0.75$ Myr for $\lambda_{\mathrm{p}} = 0.01$ due to the saturation of SSI~\citep[][]{StoneEdelman1995ApJ...454..182S}.

\begin{figure}[ht!]
\plotone{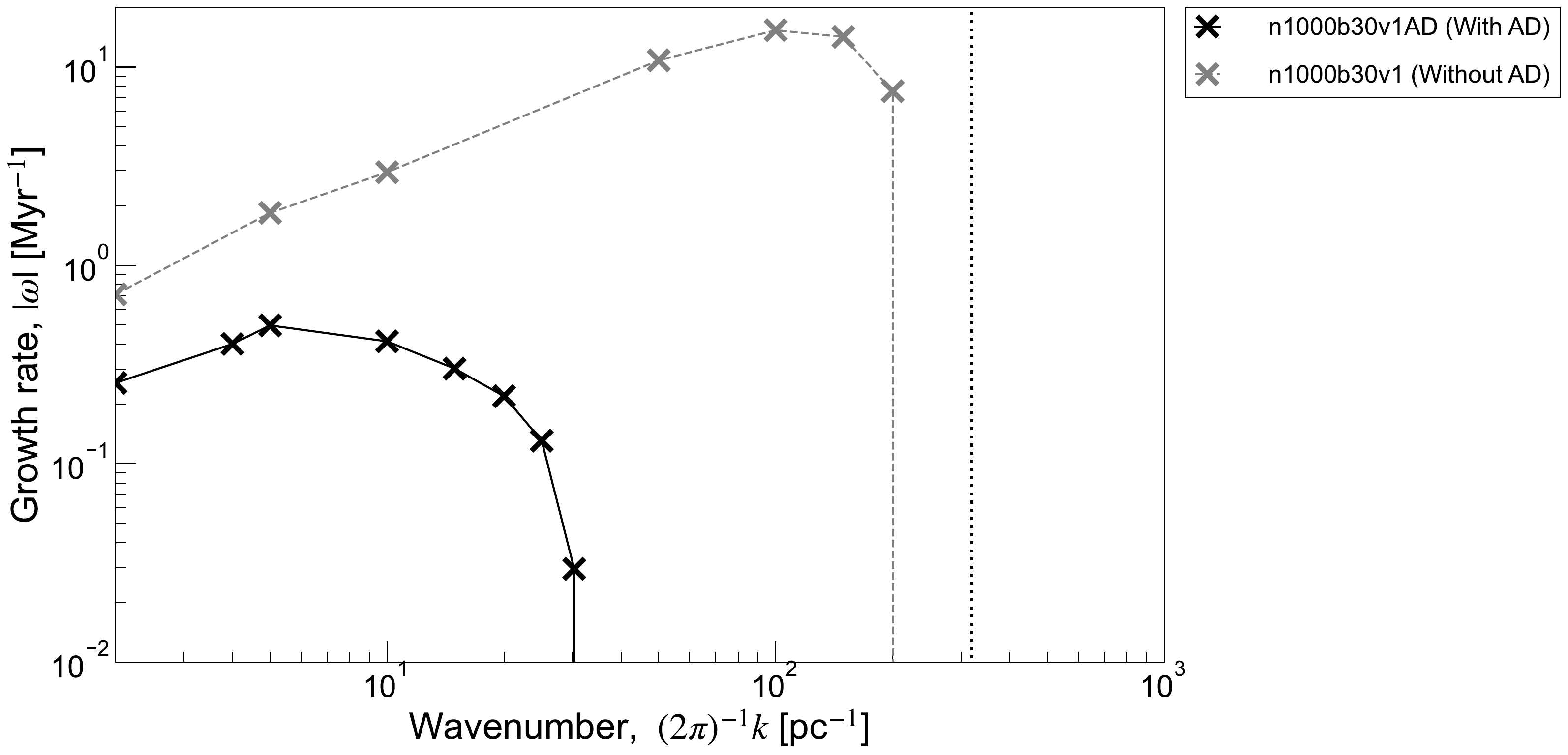}
\caption{\small{{Effect with and without ambipolar diffusion [model n1000b30v1 (gray) and n1000b30v1AD (black). The vertical dotted line represents the grid scale of $8\Delta x$.}}\label{fig:dispiso_ambi}}
\end{figure}

The slope of each line in Figure \ref{fig:byevo_iso} reflects the growth rate.
The growth rate can be measured from the slope of $\left\langle B_{y}^{2}\right\rangle$ as
\begin{equation}
\omega_{\mathrm{num}} = \left(2 \log_{10} e \right)^{-1} \frac{d}{dt} \log_{10} \frac{\left\langle B_{y}^{2}\right\rangle}{B^2_0} \simeq \left(2 \log_{10} e \right)^{-1} \frac{\Delta \log_{10} {\left\langle B_{y}^{2}\right\rangle / B^2_0}}{\Delta t}.
\end{equation}
where, $\Delta \log_{10} \left({\left\langle B_{y}^{2}\right\rangle / B^2_0}\right) / \Delta t$ denotes the gradient in the $t$--$\log_{10} \left\langle B_{y}^{2}\right\rangle/B^2_0$ plane.
Since the initial perturbation is not given as the eigen state of the SSI, the SSI does not start growing at $t=0$.
Thus we define the measuring range as [$t_0 = t_{\mathrm{start}}$ + $f t_{\mathrm{growth}}$, $t_0 + t_{\mathrm{range}}$] to observe the linear growth of SSI, where, $t_{\mathrm{growth}}\equiv 1/\omega_{\mathrm{ana}}$ is a growth timescale.
In this section, we select $t_{\mathrm{start}}=0.05$ Myr, $t_{\mathrm{range}}=0.5$ Myr, and $f=0.6$.
We show the dispersion relation for the isothermal ideal MHD case including the physical shear viscosity (model n1000b30v1, $\mathcal{R}_{\mathrm{shear}}=19.5$) as the gray cross marks in Figure \ref{fig:dispiso_ambi}.
The vertical dotted line represents the scale of $8\Delta x$.
We find that if we use $\mathcal{R}_{\mathrm{shear}}=19.5$, a physical dispersion relation is successfully obtained by suppressing the carbuncle phenomenon.

\subsection{SSI v.s. Ambipolar Diffusion}\label{subsec: SSI v.s. Ambipolar Diffusion}
\begin{figure}[ht!]
\plotone{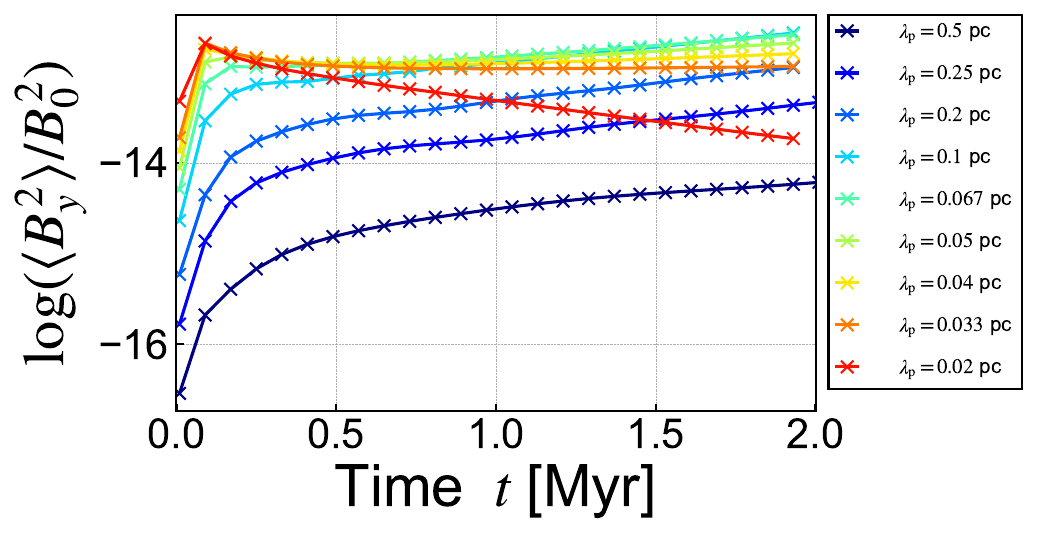}
\caption{\small{Evolution of mean value of the perturbed magnetic field for the cases with isothermal ideal MHD including ambipolar diffusion and physical shear viscosity (model n1000b30v1AD).}\label{fig:byevo_iso_ambi}}
\end{figure}

\begin{figure}[ht!]
\epsscale{0.9}
\plotone{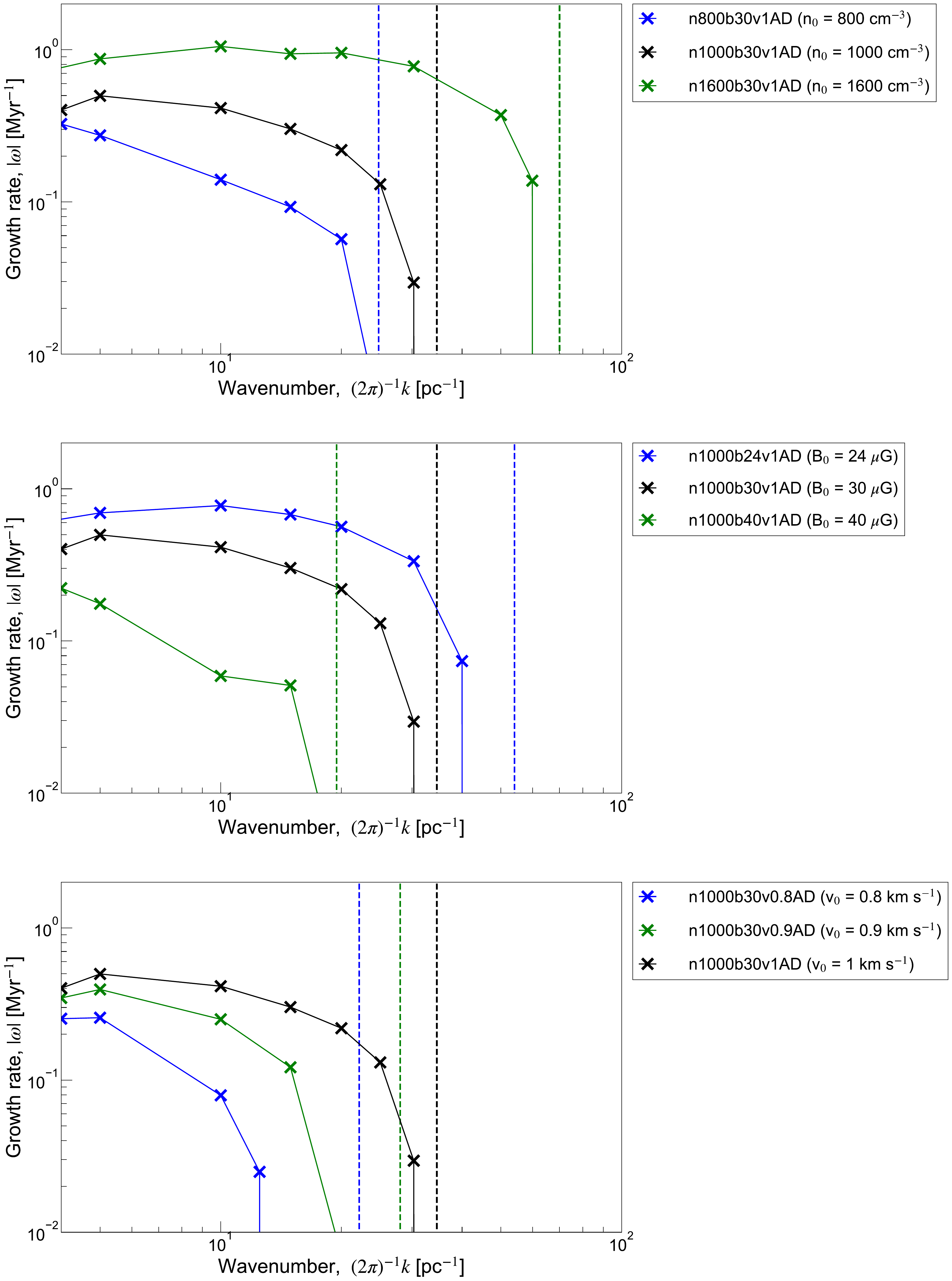}
\caption{\small{{\textit{Top panel}: Effect of varying the initial density. Blue, black, and green crosses represent the growth rate for the models n800b30v1AD, n1000b30v1AD, and n1600b30v1AD, respectively. The vertical dashed lines for each color represent the 1/($1.7\ell_{\rm AD}$) for the model corresponding to that color.
\textit{Middle panel}: Same as top panel, but for cases with varying the initial magnetic field [models n1000b24v1AD (blue), n1000b30v1AD (black), and n1000b40v1AD (green)].
\textit{Bottom panel}: Same as top panel, but for cases with varying the shock velocity [models n1000b30v1AD (black), n1000b30v0.9AD (green), and n1000b30v0.8AD (blue)].}
\label{fig:disp_pramdependence}}}
\end{figure}
We perform a similar analysis as in \S\ref{subsec: SSI in an ideal MHD case} for the simulation results including ambipolar diffusion.
In Figure \ref{fig:byevo_iso_ambi}, we show the evolution of the mean value of the perturbed magnetic field for model n1000b30v1AD.
Because the effect of ambipolar diffusion diminishes the phase speed of the Alfv\'{e}n wave, the eigen state requires more time than the ideal MHD case to develop from the given initial perturbation.
Thus, to measure the linear phase growth rate, we take longer $t_{\mathrm{start}}$ of $1.2$ Myr, and $f=0.6$ and $t_{\mathrm{range}} = 0.3$ Myr.
We show the dispersion relation for the isothermal MHD case including ambipolar diffusion (model n1000b30v1AD) as the black cross marks in Figure \ref{fig:dispiso_ambi}.
We can see the reduction of the SSI growth by the ambipolar diffusion.
We find that the most unstable scale $\ell_{\rm max} \simeq$ 0.2 pc and the damping scale $\ell_{\rm damp} \simeq$ 0.02 pc.

To investigate the parameter dependence on the most unstable scale, we perform a parameter survey for the unperturbed magnetic field strength, density, and shock velocity.
In Figure \ref{fig:disp_pramdependence}, we show the dispersion relation for models n1000b24v1AD, n1000b40v1AD, n800b30v1AD, n1600b24v1AD, and n1000b30v0.8AD.
We can confirm that the most unstable scale varies with the magnetic field density and velocity, such that a larger density/velocity corresponds to a smaller most unstable scale.
Conversely, a larger magnetic field corresponds to a larger most unstable scale.
These trends can be {understood} based on the scale of ambipolar diffusion $\ell_{\rm damp} \sim \ell_{\mathrm{AD}} \propto B_0^{2} \rho^{-3/2} v_{x0}^{-1}$ (Eq. [\ref{eqs:ambi scale}]).




\subsubsection{Three dimensional simulations}\label{subsubsec: Three dimensional simulations}
\begin{figure}[ht!]
\plotone{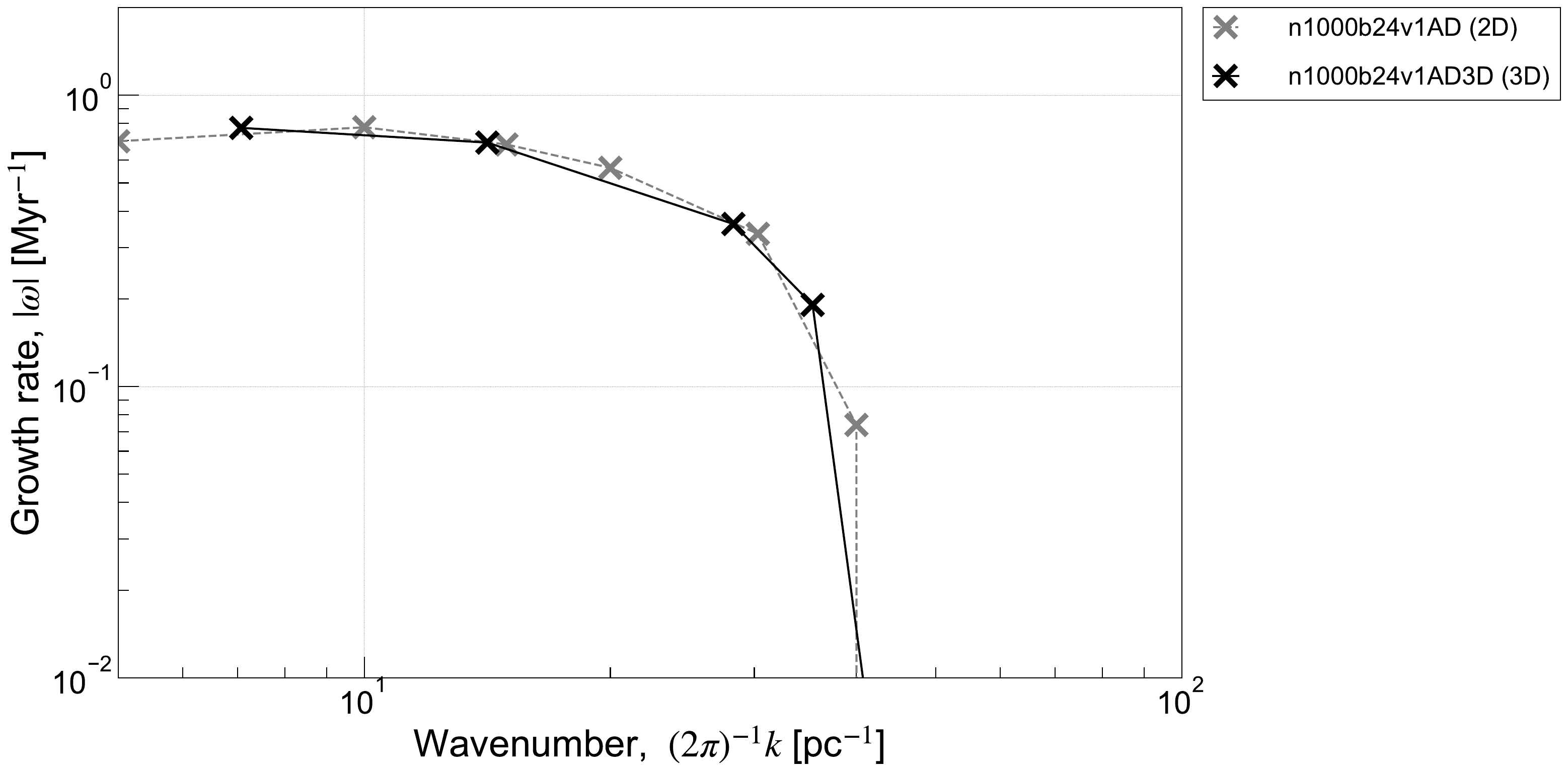}
\caption{\small{{The dependence on the number of spatial dimensions [model n1000b3024v1AD3D (black) and n1000b24v1AD (grey)].
The $k$ is defined as $\sqrt{k_y^2 + k_z^2} = 2 \sqrt{2} \pi / \lambda_p$ for the model n1000b3024v1AD3D.}
\label{fig:dispiso_ambi_sts3d}}}
\end{figure}
{We also perform three-dimensional simulations to investigate more realistic cases in molecular clouds.
Because of the higher computational cost for 3D simulations, we used the super-time stepping method for the diffusion term.
In Appendix \ref{subsubsec: Test for the Super Time Stepping method}, we show the results of the tests for the super-time stepping method (STS, \citet{Meyer2014JCoPh.257..594M}) and confirm that the results do not change even with considerable acceleration of the calculations.
In Figure \ref{fig:dispiso_ambi_sts3d}, we show the dispersion relations for models n1000b24v1AD3D (black solid line) and n1000b24v1AD (grey dashed line).
We measure the growth rate in the same way as \S \ref{subsec: SSI v.s. Ambipolar Diffusion}.
We can confirm that those two are not different regardless of dimensions.}

\section{Discussion} \label{sec: Discussion}
\begin{figure}[ht!]
\epsscale{0.8}
\plotone{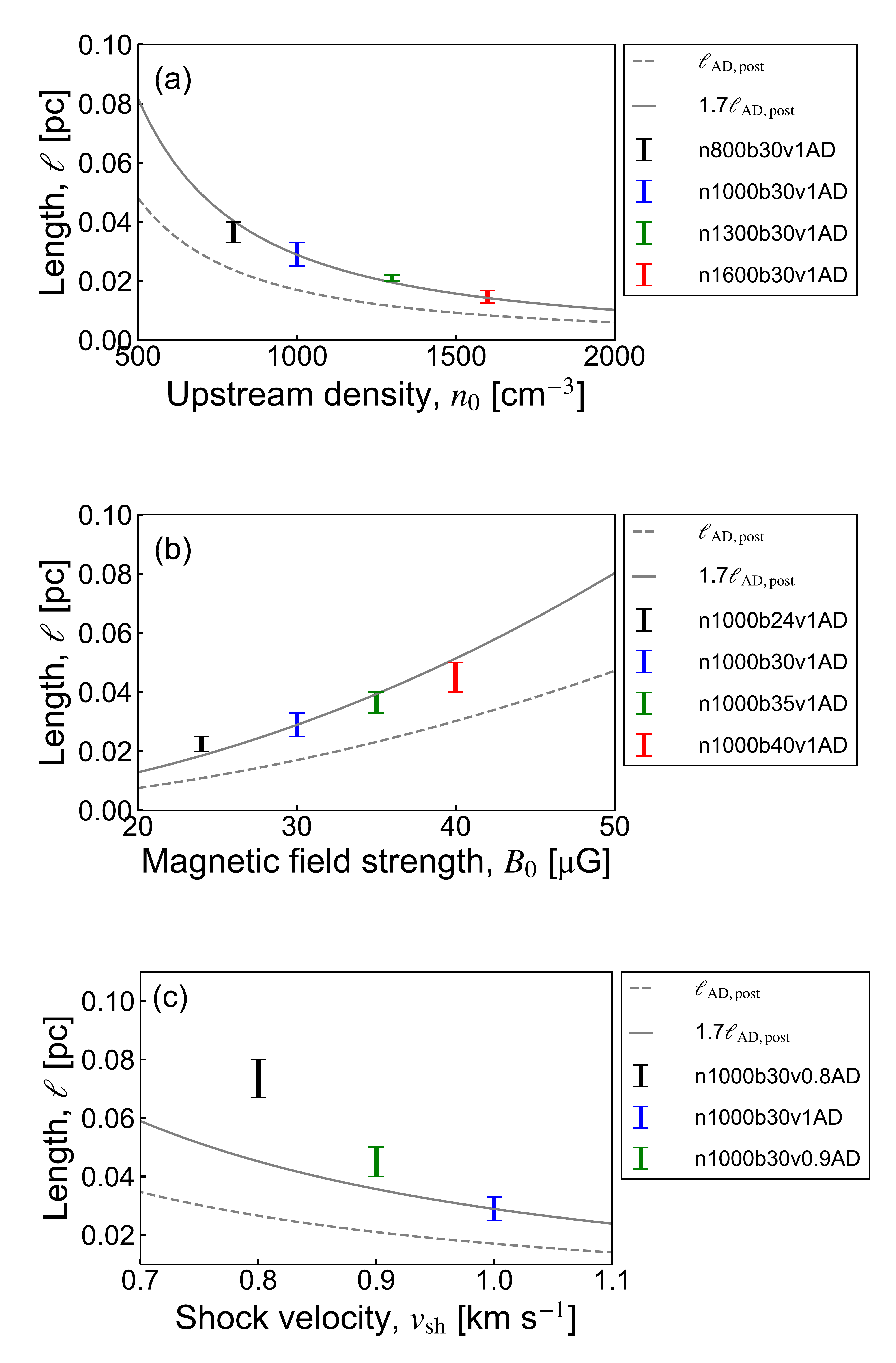}
\caption{\small{Parameter dependence of damping scale $\ell_{\rm damp}$ (solid vertical lines). Gray dashed and solid lines represent $\ell_{\rm AD,post}$ (Eq. \ref{eqs:ambi scale post}) and $1.7\ell_{\rm AD,post}$. (a) Upstream density versus damping scale. (b) Magnetic field versus damping scale. (c) Shock velocity versus damping scale.\label{fig:discussion_dampingscale}}}
\end{figure}
In \S \ref{subsec: SSI v.s. Ambipolar Diffusion}, we have stated that the most unstable scale depends on the magnetic field, density, and shock velocity.
In this section, {we discuss how the most unstable wavelength of the SSI is scaled.}
In Figure \ref{fig:discussion_dampingscale}, we show the damping length scale, which {scales} with the most unstable length, as a function of the initial density (panel a), the initial magnetic field (panel b), and the shock velocity (panel c).
The gray dotted line is the scale of ambipolar diffusion evaluated in post-shock quantities which can be written as
\begin{equation}
\ell_{\mathrm{AD,post}}
\equiv \frac{ B_1^2}{4\pi \gamma_{\rm in} C \rho_{1}^{3/2} v_1}
\simeq \frac{ B_0^2 c_{\rm s}}{4\pi \gamma_{\rm in} C \rho_{0}^{3/2} v_0^2}
= 0.017\ \mathrm{pc}\ \left(\frac{B_0}{30\ \mathrm{\mu G}}\right)^{2}\left(\frac{n_0}{10^3\ \mathrm{cm^{-3}}}\right)^{-3/2}\left(\frac{v_0}{1\ \mathrm{km/s}}\right)^{-2},
\label{eqs:ambi scale post}
\end{equation}
where, $B_1\simeq B_0$, $\rho_1 \simeq \mathcal{M}_{\rm s}^2 \rho_0$, and $v_1 \simeq \mathcal{M}_{\rm s}^{-2} v_0$ denote the post-shock magnetic field strength, density, and velocity, respectively.
The gray solid line represents $1.7\times \ell_{\mathrm{AD,post}}$.
The top and bottom edges of vertical lines denote the minimum lengths with a positive growth rate and the maximum lengths with a negative growth rate, i.e., the vertical lines show the range in which the damping scale exists.
For panels (a) and (b), the damping length scale well follows $1.7\ell_{\mathrm{AD, post}}$.
{The dependence on shock velocity deviates more strongly from the predictions of Eq. (\ref{eqs:ambi scale post})}, but the difference from $1.7\ell_{\mathrm{AD, post}}$ is within a factor of 2.
{While our estimation for the length scale of ambipolar diffusion utilized physical quantities from the downstream region, a more appropriate approach should be based on the shock transition layer where ambipolar diffusion actually works. The dependency on the magnetic field and upstream density in Eq. (\ref{eqs:ambi scale post}) remains consistent because the change of the magnetic field or upstream density does not shift the compression ratio (Fig. \ref{fig:discussion_dampingscale} a, b).
However, differently from density and magnetic field strength, the shock velocity affects the compression ratio in the isothermal case (compression ratio $r \propto \mathcal{M}^2$) that brings some error for the ambipolar diffusion scale estimation (Eq. \ref{eqs:ambi scale post}).
The density in the transition layer is lower than the downstream density, which implies that the actual length scale of ambipolar diffusion is expected to be larger than Eq. (\ref{eqs:ambi scale post}). This may account for the observed deviation between the simulation results and Eq. (\ref{eqs:ambi scale post}).}
We conclude that the characteristic damping scale of the SSI can be approximately estimated as $\ell_{\mathrm{damp}} \sim 1.7\ell_{\mathrm{AD, post}}$, and the most unstable scale as $\ell_{\mathrm{SSI}} \sim 5\ell_{\mathrm{damp}} \sim 9\ell_{\mathrm{AD,post}}$.

{The filament width is one of the significant quantities to determine the initial condition for star formation.
The critical line-mass including the support of the magnetic field depends on the filament width~\citep[][]{Tomisaka2014ApJ...785...24T}.
According to linear theory, the self-gravitational fragmentation length scale of the filament depends on the filament width~\citep[][]{Stodolkiewicz1963,InutsukaMiyama1992}.}
{\citet{Arzoumanian2011A&A...529L...6A} {found} that the characteristic width of the \textit{Herschel} Gould Belt filaments is 0.1 pc~\citep[see also][]{Juvela2012A&A...541A..12J,KochRosolowsky2015MNRAS.452.3435K}.
Remarkably, the filaments maintain their width, regardless of their line-mass exceeding 100 $M_{\rm \odot}$ pc$^{-1}$.}
{It should be noted that some studies have questioned the universality of 0.1 pc filament widths.
For example, \citet{Ossenkopf-Okada2019A&A...621A...5O} used wavelet decomposition to analyze \textit{Herschel} survey data and did not find a characteristic length scale.
{\citet{Panopoulou2022A&A...657L..13P,Panopoulou2022A&A...663C...1P} suggested that the estimated filament width appears to depend weakly on the distance to the observed cloud. }
}
{Considering only thermal support against gravity, such a high-line-mass structure cannot maintain a length scale of 0.1 pc.
Several authors studied the effects of turbulence and/or the magnetic field and have shown that sub-critical and mildly super-critical filament has a width of 0.1 pc~\citep[][]{FischeraMartin2012A&A...542A..77F,Auddy2016ApJ...831...46A,PriestleyWhitworth2022MNRAS.512.1407P,Federrath2016}, however, the reason for the constant width of filaments, especially for massive filaments, remains as a mystery.}
{\citet{Seifried2015MNRAS.452.2410S} simulated the evolution of a massive filament with a line-mass of 75 $\rm {M_{\odot}\ pc^{-1}}$ and found that filaments perpendicular to the magnetic field are thinner.
{They claimed that their very narrow filaments could be interpreted as the fibers reported by \citet{Hacar2013A&A...554A..55H}.}
}
{\citet{Smith2014MNRAS.445.2900S} performed simulations of filament formation, which indicated that the width of the massive filament is approximately 0.3 pc which is inconsistent with the constant 0.1 pc filament width proposed by \citet{Arzoumanian2011A&A...529L...6A}.
} 
{We need to understand the origin of the universal width, especially for massive filaments, by examining the detailed process of gas accretion flows onto the filament.}

\section{Summary} \label{sec: summary}

We performed the SSI simulations in molecular clouds using two-dimensional isothermal ($\gamma=1.01$) or adiabatic ($\gamma=5/3$) using Athena++.
Furthermore, we investigated the most unstable length scale of SSI in molecular clouds by ambipolar diffusion, {which provides the resolution requirement in future simulations.}
The major findings of this study are stated as follows.
\begin{enumerate}
  \item Ambipolar diffusion suppresses SSI on a small scale. We find that the most unstable scale of the order of $\sim$ 0.1 pc and the damping scale of the order of $\sim$ 0.01 pc in molecular clouds.
  \item The most unstable and damping length scales depend on the density, shock velocity, and, magnetic field strength. The scaling is roughly described by $\ell_{\rm SSI} \sim 5\ell_{\rm damp} \sim 9\ell_{\mathrm{AD,post}} \propto B_0^{2} \rho^{-3/2} v_{x0}^{-2}$.
\end{enumerate}

{In a follow-up paper (Paper II)}, we will show that a natural extension of this work leads to a better understanding of the filament evolutionary process.
In realistic situations, the filaments are bound by two shocks. Because the separation of these two shocks is narrow and the two shock {surfaces} share the magnetic field lines that thread them, we can expect that the two shocks dynamically influence each other. In Paper II, we will discuss the effect of {two interacting shocks} and demonstrate that the SSI creates inhomogeneous postshock flows and can provide additional dynamical pressure to the filament.

\begin{acknowledgments}
We thank K. Tomida for the fruitful discussions. The numerical computations were carried out on the XC50 system at the Center for Computational Astrophysics (CfCA) of the National Astronomical Observatory of Japan. This work is supported by Grant-in-aids from the Ministry of Education, Culture, Sports, Science, and Technology (MEXT) of Japan (JP22J15861, 18H05436).
\end{acknowledgments}

\vspace{5mm}

\appendix

\begin{table*}
\caption{Model parameters.\label{tab:modelparameters2}}
 \centering
  \begin{tabular}{cccccccccc}
   \hline
  Model Name  & $n_0$  & $B_{0}$ & $v_{\mathrm{x0}}$ & $\gamma$ & $\mathcal{R}_{\mathrm{shear}}$ & ambipolar & Solver & STS & Dimension \\
  & [cm$^{-3}$] & [$\mu$G] & [km s$^{-1}$] & & $\equiv v_{\mathrm{x0}} \Delta x/{\nu}$ & diffusion & & &\\
   \hline \hline
    n1000b30v1a$\mathcal{R}\infty$   & 1000 & 30  & 1.0  & 5/3  & $\infty$ & No & Roe & No & 2D \\
    n1000b30v1a$\mathcal{R}$16.3     & 1000 & 30  & 1.0  & 5/3  & 16.3     & No & Roe & No & 2D  \\
    n1000b30v1a$\mathcal{R}$9.8      & 1000 & 30  & 1.0  & 5/3  & 9.8      & No & Roe & No & 2D  \\
    n1000b30v1a$\mathcal{R}\infty$E  & 1000 & 30  & 1.0  & 5/3  & $\infty$ & No & HLLE & No & 2D  \\
    n1000b30v1a$\mathcal{R}\infty$D  & 1000 & 30  & 1.0  & 5/3  & $\infty$ & No & HLLD & No & 2D  \\
    n1000b30v1a$\mathcal{R}$16.3D    & 1000 & 30  & 1.0  & 5/3  & 16.3     & No & HLLD & No & 2D  \\
    n1000b30v1a$\mathcal{R}$9.8D     & 1000 & 30  & 1.0  & 5/3  & 9.8      & No & HLLD & No & 2D  \\
    n1000b30v1a$\mathcal{R}\infty$LD & 1000 & 30  & 1.0  & 5/3  & $\infty$ & No & LHLLD & No & 2D  \\
    n1000b30v1a$\mathcal{R}$16.3LD   & 1000 & 30  & 1.0  & 5/3  & 16.3     & No & LHLLD & No & 2D  \\
    n1000b30v1a$\mathcal{R}$9.8LD    & 1000 & 30  & 1.0  & 5/3  & 9.8      & No & LHLLD & No & 2D  \\
    n1000b30v1.3a$\mathcal{R}\infty$D & 1000 & 30  & 1.3  & 5/3  & $\infty$   & No & HLLD & No & 2D   \\
    n1000b30v1.3a$\mathcal{R}\infty$LD & 1000 & 30  & 1.3  & 5/3  & $\infty$   & No & LHLLD & No & 2D  \\
    n1000b30v1AD-STS& 1000 & 30  & 1.0  & 1.01 & 19.5 & Yes & Roe & Yes & 2D  \\
   \hline
  \end{tabular}
\end{table*}

\section{Examination of Numerical Scheme and Physical Viscosity} \label{subsubsec: Selection of MHD Solver}
\begin{figure}[ht!]
\plotone{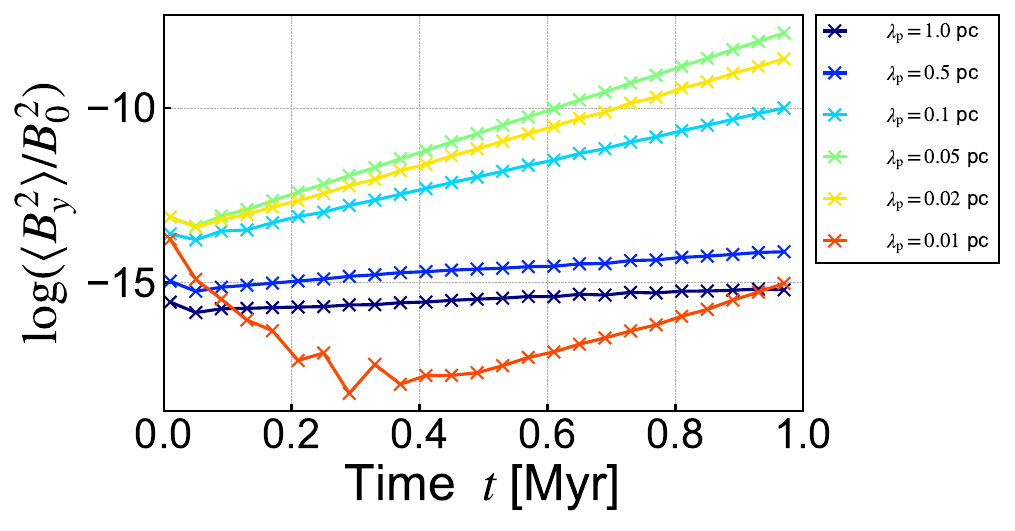}
\caption{Evolution of mean value of the perturbed magnetic field for the case with adiabatic ideal MHD including physical shear viscosity (model n1000b30v1$\mathcal{R}$9.8).\label{fig:byevo_adi}
}
\end{figure}
We perform the simulations with $\gamma = 5/3$ and compare the results with the dispersion relation (Eq. [\ref{eqs:ana disp}]), based on which we can check whether the selected numerical scheme appropriately reproduces at least the regime of the linear instability.
We select $L_{\mathrm{box}}=1$ pc.
{In Table \ref{tab:modelparameters2}, we show model parameters to test.
Each model has a unique name, the rule of name is the basically same as Table 1.
Models with $\gamma$ = 5/3 are additional denoted as ``a”, followed by the Reynolds number of the physical shear viscosity (``9.8,” ``16.3,” ``19.5," and ``$\infty$").
Models with various MHD solvers are additionally denoted as ``E," ``D," and ``LD" corresponding to HLLE, HLLD, and LHLLD, respectively.
For simulations using the super-time stepping method, we added the notation ``STS."}
For $\lambda_{\mathrm{p}}=0.01$ pc, $\left\langle B_{y}^{2}\right\rangle/B^2_0$ decreases until $\sim$ 0.3 Myr and a larger scale noise caused by numerical error starts to grow after $t \sim$ 0.3 Myr, which varies from the growth of $\lambda_{\mathrm{p}}=0.01$ pc mode.
We measure the growth rate in the same way as \S \ref{subsec: SSI in an ideal MHD case} and select $t_{\mathrm{ini}}=0.05$ Myr, $t_{\mathrm{range}}=0.2$ Myr and $f=0.4$.

\begin{figure}[ht!]
\plotone{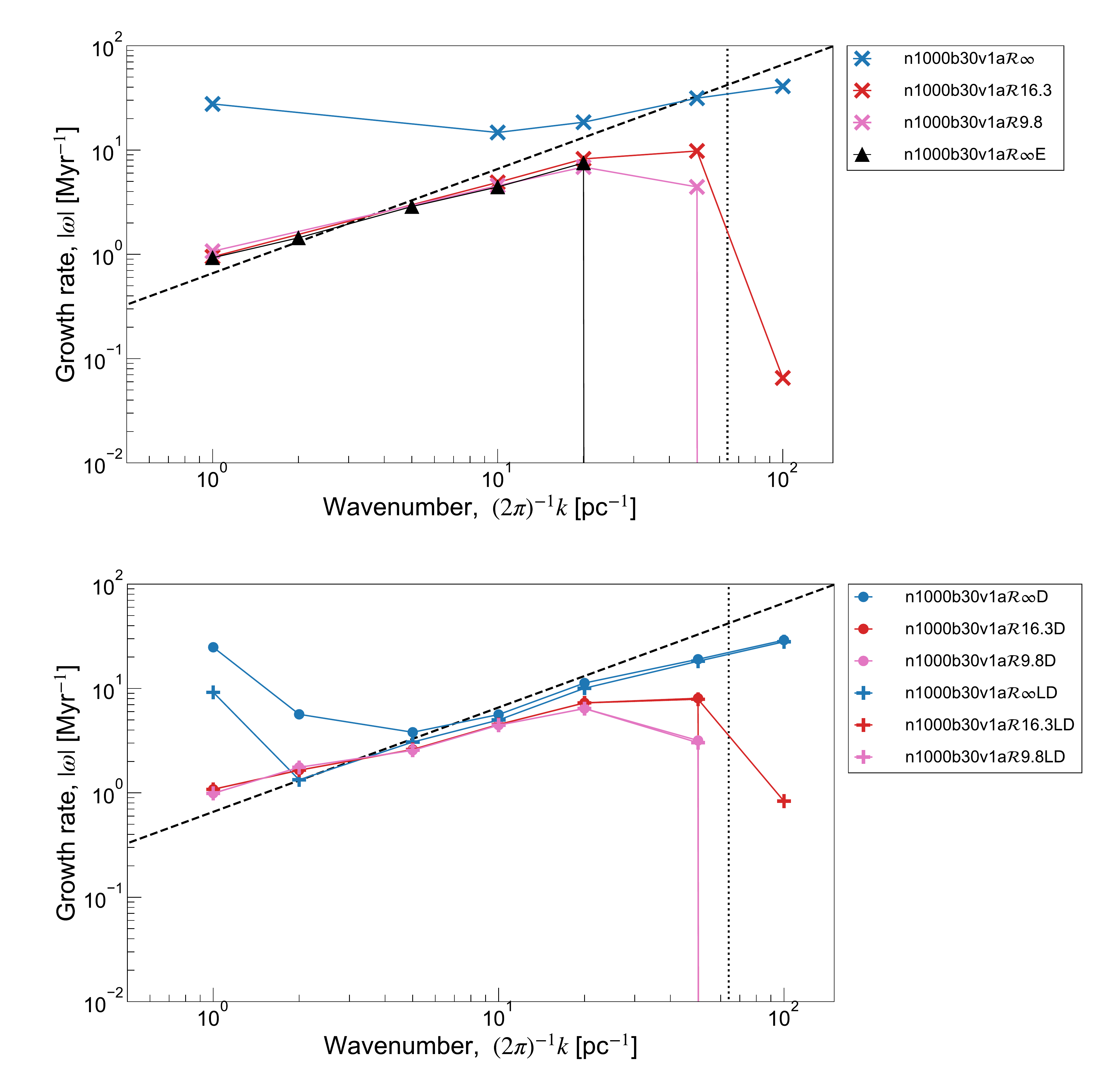}
\caption{\label{fig:dispadi}
\textit{Top panel}: Dispersion relations for cases with Roe solver (cross marker, models n1000b30v1a$\mathcal{R}\infty$, n1000b30v1a$\mathcal{R}16.3$, and n1000b30v1a$\mathcal{R}9.8$) and HLLE solver (triangle marker, model n1000b30v1a$\mathcal{R}\infty$E) solvers. Black and blue lines denote the non-viscous results. Red and pink lines represent the result with $\mathcal{R}_{\mathrm{shear}}=16.3$ and  $\mathcal{R}_{\mathrm{shear}}=9.8$.
\textit{Bottom panel}: Same as top panel, but for cases with HLLD solver (circle marker, models n1000b30v1a$\mathcal{R}\infty$D, n1000b30v1a$\mathcal{R}16.3$D, and n1000b30v1a$\mathcal{R}9.8$D) and LHLLD solver (plus marker, models n1000b30v1a$\mathcal{R}\infty$LD, n1000b30v1a$\mathcal{R}16.3$LD, and n1000b30v1a$\mathcal{R}9.8$LD). 
}
\end{figure}

The HLLD and LHLLD Riemann solvers are robust and high-resolution MHD solvers, but it does not take into account the slow mode characteristics in the Riemann problem, and it is necessary to test the optimal solver for solving SSI.
We perform simulations using HLLD, LHLLD, HLLE, and Roe solvers to decide which solver is the best for solving SSI.
If physical shear viscosity is not included, the measurement of the growth rate fails due to the carbuncle phenomenon.
We can prevent the carbuncle phenomenon by introducing small physical shear viscosity.
In addition to the carbuncle phenomenon, grid noise, which is numerically induced noise and whose scale is different from a given $\lambda_{\rm p}$, potentially becomes the seed of SSI.
Since the small scale fluctuation grows faster, the growth of SSI seeded by the grid noise can contaminate growth rate measuring after a long time integration (e.g., see the red line in Figure [\ref{fig:byevo_adi}]), thus we avoid measuring the growth of grid noise by adjusting the $t_{\rm range}$.

The top panel in Figure \ref{fig:dispadi} shows the dispersion relation for adiabatic cases calculated using the HLLE or Roe solvers (models n1000b30v1a$\mathcal{R}\infty$E, n1000b30v1a$\mathcal{R}\infty$, n1000b30v1a$\mathcal{R}16.3$, and n1000b30v1a$\mathcal{R}9.8$).
The dashed line represents Eq. (\ref{eqs:ana disp}), and the vertical dotted line represents the scale of $\lambda = 8\Delta x$.
We can see that the grid scale structure created by the carbuncle phenomenon is suppressed by introducing the physical shear viscosity (compare the cross marks).
For simulations employing HLLE solver, we do not introduce the physical shear viscosity because of its very diffusive nature, however the diffusive nature attenuates the SSI at a scale larger than the Roe cases with physical viscosity (see filled triangles).
In the results with the Roe solver without introducing the physical shear viscosity (blue cross marks), we cannot measure growth rates at long wavelength regimes owing to the contamination by the carbuncle phenomenon.
For model n1000b30v1$\mathcal{R}$16.3 (red crosses), the carbuncle phenomenon still appears at $\lambda_{\rm p}$ = 0.02 pc (see the right edge of red curve).
Using the Roe method with $\mathcal{R}_{\mathrm{shear}}=9.8$ (pink crosses), the SSI growth rate can be measured with high resolution (down to $\sim$ 0.02 pc) as well as preventing the carbuncle phenomenon.

The bottom panel of Figure \ref{fig:dispadi} shows the same as the top but results using HLLD or LHLLD solvers (models n1000b30v1a$\mathcal{R}\infty$D, n1000b30v1a$\mathcal{R}16.3$D, n1000b30v1a$\mathcal{R}9.8$D, n1000b30v1a$\mathcal{R}\infty$LD, n1000b30v1a$\mathcal{R}16.3$LD, and n1000b30v1a$\mathcal{R}9.8$LD).
The simulation using HLLD suffers from the carbuncle phenomenon.
The LHLLD scheme is designed to alleviate the carbuncle phenomenon, but after a long time integration, the growth of grid noise appears.
Both schemes provide similar results, but we fail to measure for $\lambda_{\mathrm{p}}\ >$ 1.0 pc modes if we do not involve physical shear viscosity by the effects of the carbuncle phenomenon due to slower growth of the SSI than the carbuncle phenomenon (see filled blue circles and blue plus marks).
It should be mentioned that the results using LHLLD are closer to the approximated analytical solution than those using HLLD at $\lambda_{\mathrm{p}} \leq$ 0.5 pc.
The simulations using either HLLD or LHLLD successfully reproduce the growth of SSI for $\mathcal{R}_{\mathrm{shear}}=9.8$ (see pink circles and plus marks).

In a conclusion, we find that the HLLD, LHLLD, and Roe solvers with adjusted physical shear viscosity can correctly calculate the growth rate of SSI over a wide scale range.
We can use any of the HLLD, LHLLD, and Roe solvers to measure the linear growth rate, but in the following sections, we use the Roe solver that shows because of its more numerically stable features in nonlinear regimes (see Appendix \ref{subsubsec: Unphysical Numerical Explosion}).

\section{Unphysical Numerical Explosion in HLLD/LHLLD} \label{subsubsec: Unphysical Numerical Explosion}
\begin{figure}[ht!]
\epsscale{0.9}
\plotone{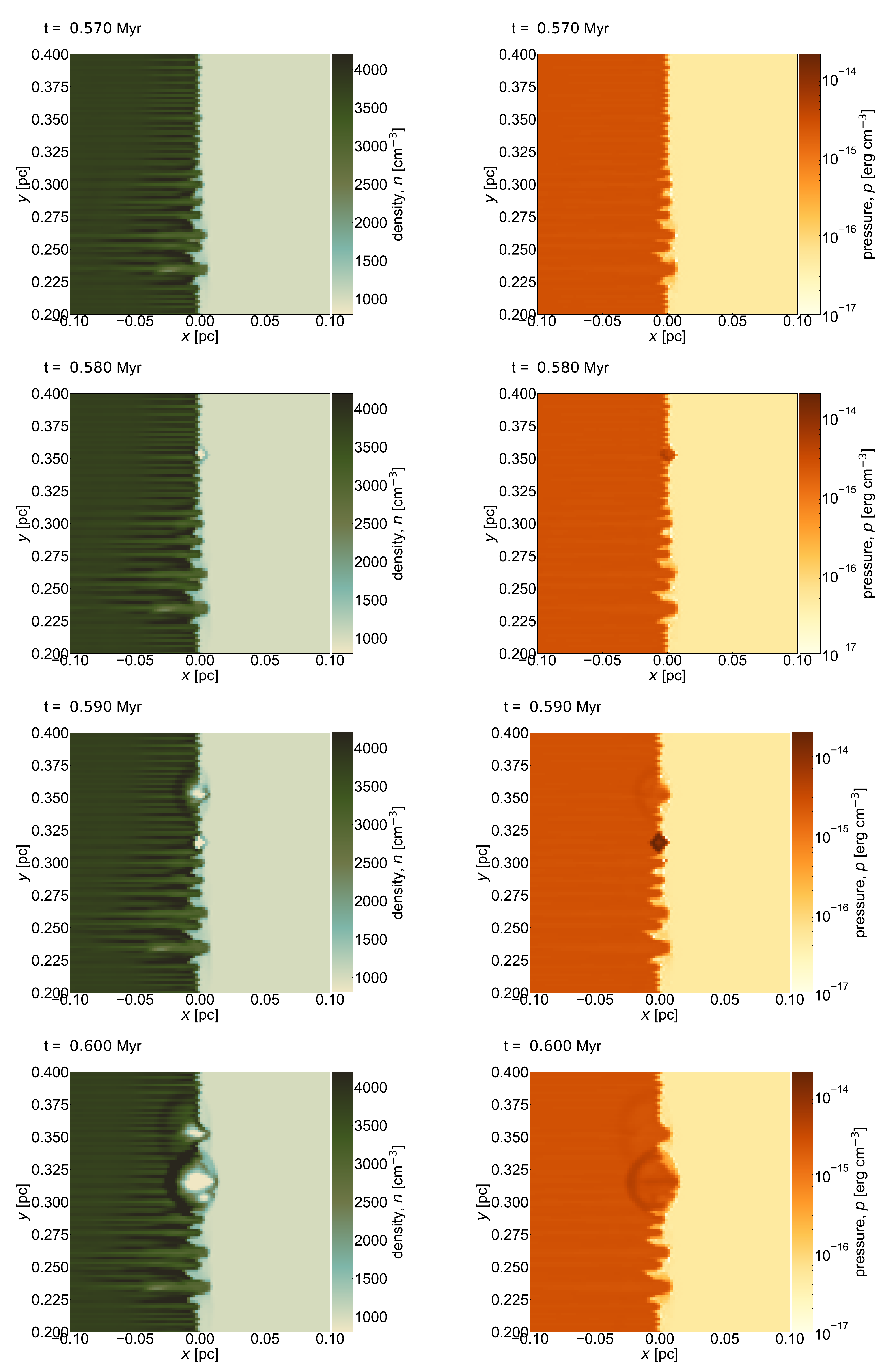}
    \caption{\small{Density (left row) and pressure (right row) maps in the result of model n1000b30v1.3a$\mathcal{R}\infty$D at time $t$ = 0.57, 0.58, 0.59, and 0.60 Myr (from top to bottom).
    }}
    \label{fig:exphlld}
\end{figure}
\begin{figure}[ht!]
\epsscale{0.9}
\plotone{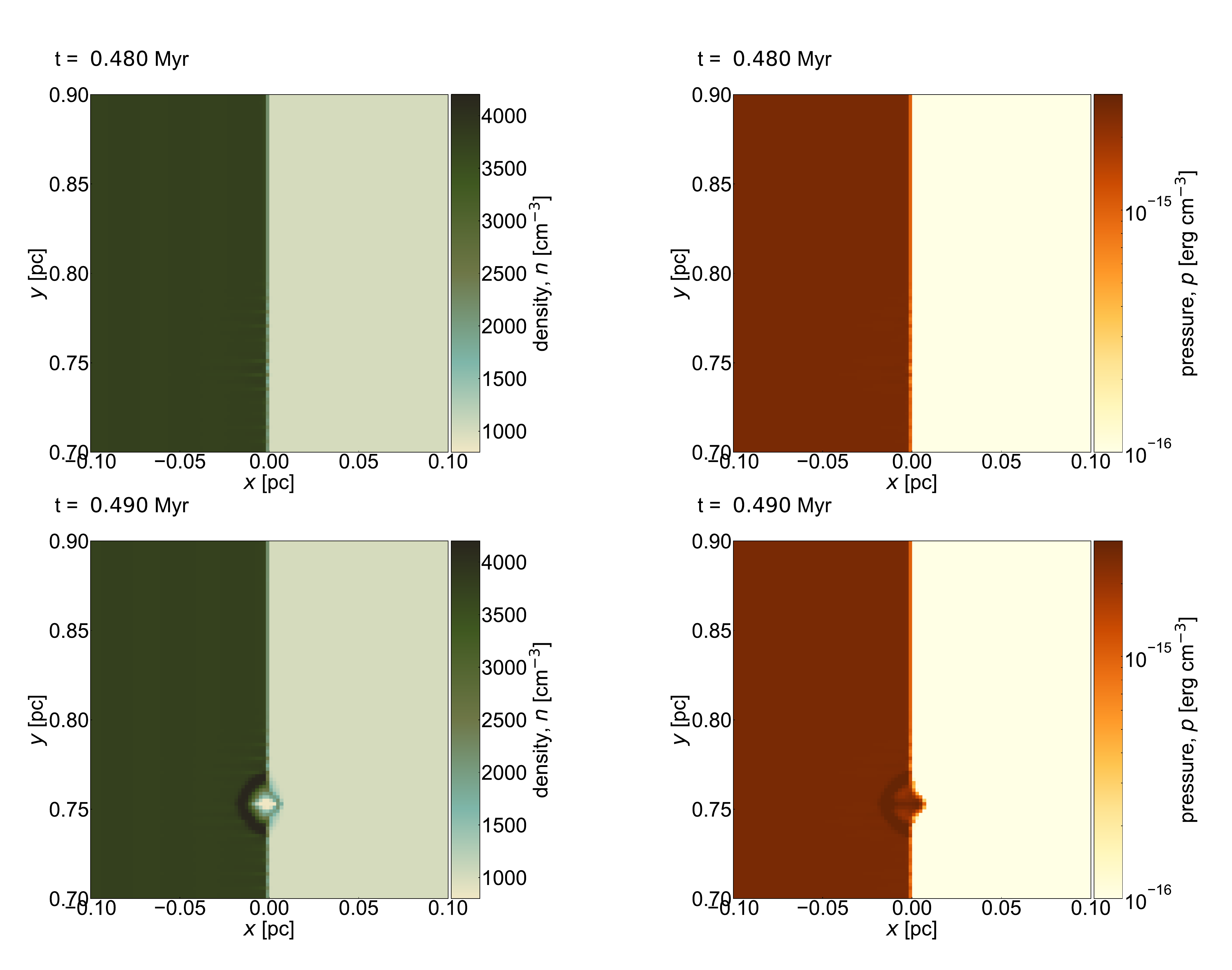}
    \caption{\small{Density (left row) and pressure (right row) maps in the result of model n1000b30v1.3a$\mathcal{R}\infty$LD at time $t$ = 0.57, 0.58, 0.59, and 0.60 Myr (from top to bottom).
    }}
    \label{fig:explhlld}
\end{figure}
In Figure \ref{fig:exphlld} and \ref{fig:explhlld}, we show snapshots of the density and pressure map of models n1000b30v1.3a$\mathcal{R}\infty$D and n1000b30v1.3a$\mathcal{R}\infty$LD, respectively.
For long term simulations using HLLD/LHLLD without the physical shear viscosity, the numerical errors around the shock front cause unphysical numerical explosions, which do not appear for simulations with Roe scheme.
This numerical problem occurs if the denominator $\rho_{\alpha}\left(S_{\alpha}-u_{\alpha}\right)\left(S_{\alpha}-S_{M}\right)-B_{x}^{2}$ in Eq. (44)-(47) of \citet[][]{MiyoshiKusano2005JCoPh.208..315M} is close to zero.
The latest version of Athena++ has been designed to prevent this issue to some extent, but it cannot prevent the unphysical explosion under the initial conditions dealt with in this study.
Although such a numerical effect can be quenched by physical viscosity, we select Roe solver with a physical shear viscosity to ensure safe long-term integration.



{
\section{Selection for a parameter in Super Time Stepping method} \label{subsubsec: Test for the Super Time Stepping method}
\begin{figure}[ht!]
\plotone{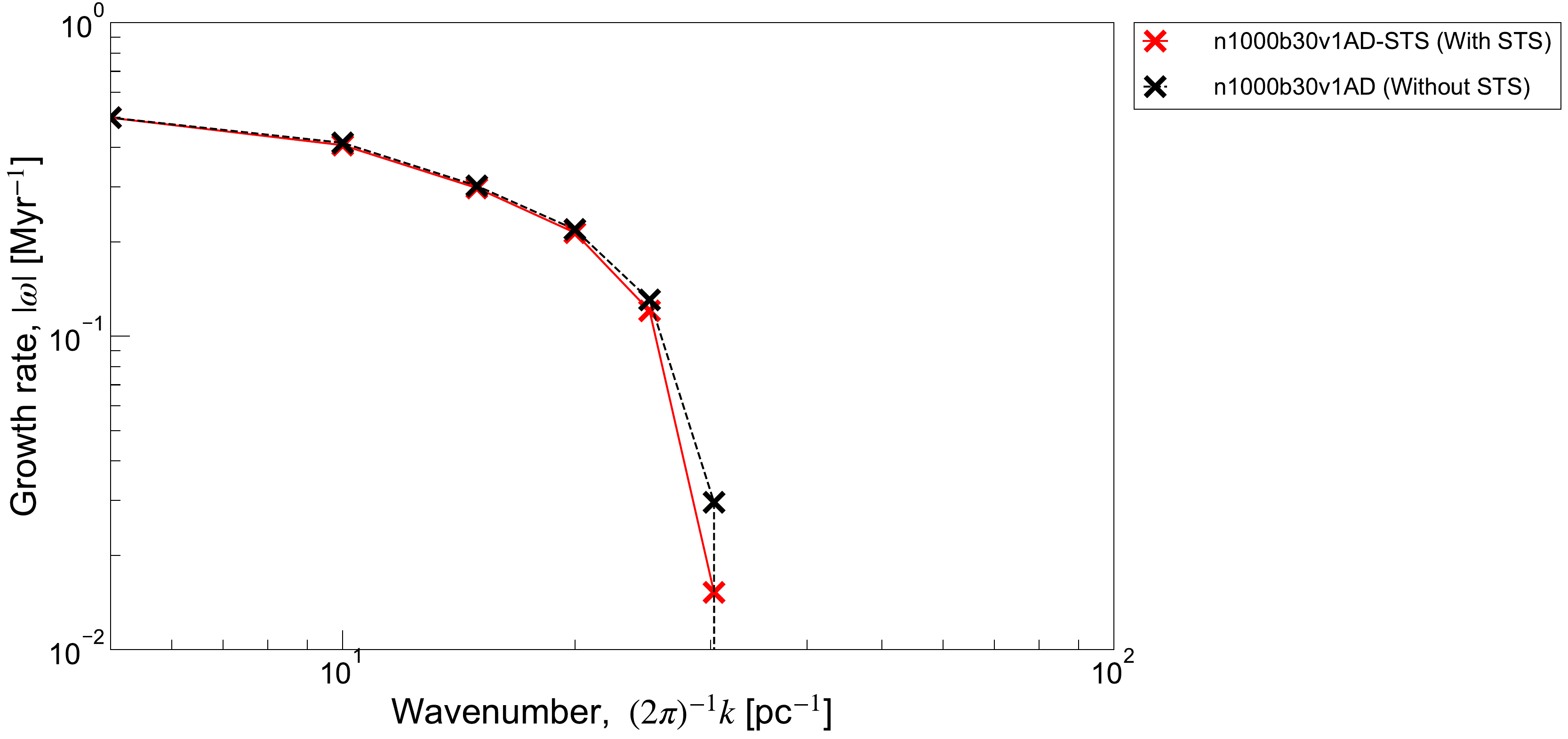}
\caption{\small{{Effect with and without the super time stepping method [model n1000b30v1AD (black) and n1000b30v1AD-STS (red)].}
    \label{fig:dispiso_ambi_sts}}}
\end{figure}
Three-dimensional simulations including ambipolar diffusion are computationally expensive and take too long to obtain results.
Thus, we use the super time stepping method~\citep{Meyer2014JCoPh.257..594M}.
The time stepping under the CFL condition is based on the condition that the solution is stable at the next time step i.e., no unphysical behavior or oscillations.
The super time-stepping method relaxes this restriction and provides a stable solution with a time step much larger than the one determined by the CFL condition.
A parameter of the super time stepping method is the maximum time step ratio $\max \left(dt/dt_{\rm parabolic} \right)$ which is the limit time step if the ratio of overall time step $dt$ calculated by fluid equations to time step $dt_{\rm parabolic}$ calculated by a diffusion equation exceeds this value.
It is necessary to test that the time integration using the super time stepping method is not significantly different from the time integration without it.
We test the super time stepping method with the same initial condition with model n1000b30v1AD and $\lambda_{\rm p}$ = 0.05 pc, and we perform a simulation with $\max \left(dt/dt_{\rm parabolic} \right) = 1000$ as model n1000b30v1AD-STS.
Figure \ref{fig:dispiso_ambi_sts} shows the dispersion relations for model n1000b30v1AD (black) and n1000b30v1AD-STS (red).
We can confirm that the results do not change significantly even {when} $\max \left(dt/dt_{\rm parabolic} \right) = 1000$.
}


\bibliography{ms}{}
\bibliographystyle{aasjournal}



\end{document}